\documentclass[aps,prb,twocolumn,showpacs]{revtex4}
\usepackage{bm}
\usepackage{amsmath}
\usepackage{graphicx}

\begin{document}


\title{Self-field effects upon the critical current density of flat
superconducting strips}

\author{Ali A. Babaei Brojeny}
\affiliation{%
 Department of Physics, Isfahan University of Technology,
Isfahan 84154, Iran, and Department of Physics and Astronomy,
  Iowa State University, Ames, Iowa, 50011--3160 }
\author{John R.\ Clem}
\affiliation{%
  Ames Laboratory and Department of Physics and Astronomy,\\
  Iowa State University, Ames, Iowa, 50011--3160 }

\date{\today}

\begin{abstract}
We develop a general theory to account self-consistently for self-field effects
upon the average transport critical current density $J_c$ of a flat type-II
superconducting strip in the mixed state when the bulk pinning is characterized
by a field-dependent depinning critical current density
$J_p(\bm B)$, where $\bm B$ is the local magnetic flux density. 
We first consider the possibility  of both bulk and edge-pinning  contributions
but conclude that bulk
pinning dominates over geometrical edge-barrier effects in state-of-the-art
YBCO films and prototype second-generation coated conductors.    
We apply our theory using the Kim model,
$J_{pK}(B) =  J_{pK}(0)/(1+|B|/B_0),$ as an example. We
calculate
$J_c(B_a)$ as a function of a perpendicular applied magnetic induction $B_a$ and
show how 
$J_c(B_a)$ is related to $J_{pK}(B)$.  We find that $J_c(B_a)$ is
very nearly equal to $J_{pK}(B_a)$  when $B_a \ge B_a^*$, where
$B_a^*$ is the  value of $B_a$ that makes the net flux density zero at the
strip's edge.  However, $J_c(B_a)$ is  suppressed relative to
$J_{pK}(B_a)$ at low fields when $B_a < B_a^*$, with the largest suppression
occurring when
$B_a^*/B_0$ is of order unity or larger.
\end{abstract}

\maketitle

\section{Introduction} 

One of the most important physical properties characterizing a type-II
superconductor is its transport critical current $I_c$.  
When the current $I$ in a
superconducting strip exceeds
$I_c$, a voltage appears along the length; this voltage is due to
the motion of vortices or antivortices across the strip.
Applying a  magnetic induction $\bm {B}_a$ perpendicular to the current
generally decreases $I_c$, and another important physical property is the
strength of the field dependence of 
$I_c(\bm{B}_a)$ vs
$\bm {B}_a$. 
For applications it is of course desirable that this field dependence be as weak
as possible, so that $I_c$ remains large in high fields.
In low applied fields, the self-field generated by the current
through the superconductor  also suppresses the critical
current.
It is the purpose of this paper to
present a method by which such self-field effects can be analyzed and calculated.

\begin{figure}
\includegraphics[width=8cm]{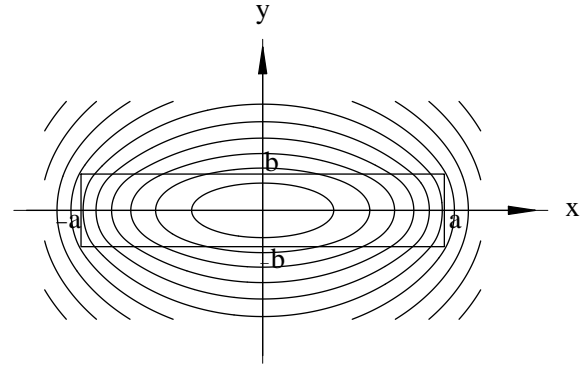}
\caption{Superconducting strip of rectangular cross section considered in this
paper.  Also shown are contours of constant vector potential $\bm A$, which
correspond to field lines of $\bm B$, when the strip carries a uniform current
density.}
\label{Fig1}
\end{figure}

The geometry of interest here is that of  a long type-II superconducting strip of
rectangular cross section  sketched in Fig. 1.  The strip is centered on the $z$
axis and is of width
$2a$ along the $x$ direction and thickness $2b$ along the $y$ direction.
For simplicity we assume that the  thickness is somewhat larger than
the London weak-field penetration depth $\lambda$, such that (a) the sample
dimensions determine the vortex nucleation conditions and (b) we can use  
macroscopic thermodynamics, which involve the magnetic induction  $\bm {B}$, a
local average of the microscopic magnetic induction over a few intervortex
spacings, and the corresponding thermodynamic magnetic field $\bm {H} =
\bm{\nabla_B}F(\bm {B})$, where $F(\bm {B})$ is the Helmholtz free energy
density in the mixed state.\cite{Campbell72}  When a magnetic
induction $\bm {B}_a$ is applied perpendicular to the $z$ axis,   $\bm {B}$ and
$\bm {H}$ have only $x$ and $y$ components, which depend only on the
coordinates $x$ and $y$.  We focus our attention on dimensions of
greatest interest to those involved in research and development on
second-generation coated conductors [highly oriented YBCO
(YBa$_2$Cu$_3$O$_{7-\delta}$) films on textured substrates]: YBCO thicknesses in
the range 0.2-4 $\mu$m and widths in the range 0.1-10 mm, such that $b \ll a$. 

Our paper is organized as follows.  In Sec.\ II, we discuss the complications
arising when both bulk and edge pinning are present, but we show that bulk
pinning dominates over geometrical edge-barrier effects in state-of-the-art
YBCO films and prototype second-generation coated conductors.  In Sec.\ III, we
discuss the general problem of how to calculate the average critical current
density $J_c({\bm B}_a)$ in an applied magnetic induction ${\bm B}_a$  when the
local depinning critical current density
$J_p$ depends upon the local magnetic induction
$\bm{B}(x,y) = \hat x B_x(x,y) + \hat y B_y(x,y)$.  In Sec.\ III A, we give some
basic equations, and in Sec.\ III B, we outline, but do not apply, a general
method requiring a two-dimensional (2D) grid.  In Sec.\ III C, we develop 
a general method
for calculating
$J_c(B_a)$ that uses only a 1D grid but requires the integral of
$1/J_p(B_x,B_y)$ with respect to
$B_x$.  This method accounts for the spatial variation of $\bm B(x,y)$ and
$J_{p}(\bm B)$ over the entire cross section of the film.
In Sec.\ III D, we apply this method for 
perpendicular applied fields $\bm B_a = \hat y B_a$ using the Kim
model,\cite{Kim63}
$J_{pK}(B) = J_{pK}(0)/(1+|B|/B_0)$, which assumes that $J_p$ depends only upon
the magnitude of $\bm B(x,y)$.
We calculate $J_c(B_a)$ vs $B_a$ and the field and current distributions across
the strip. We briefly summarize our findings in Sec.\ IV. 

\section{Bulk vs edge pinning}

According to critical-state theory in the mixed state of
type-II superconductors,\cite{Campbell72}  at the
critical current of a superconducting strip
the magnitude of the current density 
$\bm{J}(x,y) = J_z(x,y) \hat{z} = \bm{\nabla \times H}(x,y)$ in a region
containing vortices is locally equal to the critical depinning current density
$J_p(\bm{B})$.  
$J_p(\bm{B})$ characterizes the bulk pinning force per unit length $f_p$ on a
vortex but is defined to have dimensions of current density: $f_p = J_p \phi_0,$ 
where $\phi_0 = h/2e$ is the superconducting flux quantum.
On the other hand, the actual current density (averaged over a few
intervortex spacings) is $\bm{j}(x,y) = j_z(x,y) \hat{z} =\bm{\nabla \times
B}(x,y)/\mu_0$, and it is this current density that enters into the Biot-Savart
expression from which the self-field contribution to 
$\bm{B}$ can be calculated throughout all space.

Depending upon the magnitude of $J_p$, the magnetic-flux-density and
current-density distributions in a superconducting strip can be surprising
complicated because of the influence of geometrical edge barriers,
Bean-Livingston surface barriers,  or surface-barrier
effects.\cite{Bean73,Clem74,Provost74,Fortini76,
Clem79,Buzdin92,Burlachkov93,Indenbom94,Schuster94,Zeldov94a,Benkraouda96,
Doyle97,Fuchs98a,
Fuchs98b,Benkraouda98,Doyle00,Maksimov02,Elistratov02} 

For small $J_p$, the
self-field critical current of a superconducting strip is determined entirely by
the geometrical edge barrier.  For a sample in the Meissner state subjected to
an increasing current $I$ in the $z$ direction in the absence of a
perpendicular applied field, the current density in the strip averaged over the
thickness is\cite{Benkraouda98,Clem73}
\begin{equation}
j_z(x) = \frac{I}{2 \pi b \sqrt{a^2-x^2}},
\label{jzMeissner}
\end{equation}
and the $y$ component of the self-field  in the plane $y= 0$
is\cite{Clem73,Benkraouda98} 
\begin{eqnarray}
H_y(x,0)& =& 0, |x| < a, \\
&= &\frac{ x I}{2 \pi |x| \sqrt{x^2-a^2}}, |x| > a, 
\label{HyMeissner}
\end{eqnarray}
to good approximation  except within about $b$ of the edges at $x
=\pm a$.  
Irreversible vortex penetration occurs and the
critical current is reached when $|H_y(\pm a,0)|$ first reaches
the barrier-penetration field $H_s$.  In the absence of  a Bean-Livingston
barrier, $H_s = H_{c1}$, the lower critical field, but in the presence of a
Bean-Livingston barrier, $H_s$ may be much larger, even approaching the bulk
thermodynamic critical field $H_c$.\cite{Bean73,Kramer73,Fink67,Christiansen68,
Fink69} However, the value of
$H_s$ and the process by which irreversible penetration first occurs depend
sensitively upon the local structure of the
edge.\cite{Burlachkov93,Sun94,Schuster94,Zeldov94a,Benkraouda96,Benkraouda98,Morozov97,Doyle97,
Labusch97,Brandt99a,Brandt00a,Doyle00,Vodolazov03} 
The edge-barrier critical current in zero external magnetic field
$I_{s0}$ can be estimated by evaluating Eq.\ (\ref{HyMeissner}) at
$ x \approx a+b/2$, which leads to  the result  $I_{s0} = 2 \pi \sqrt{ab} H_s$
obtained in Ref.\
\onlinecite{Benkraouda98} for films with rounded
edges.\cite{Comment}  Note that when $I = I_{s0}$, the magnitude of the
self-field in the middle of the top and bottom surfaces is $|H_x(0,\pm b)| =
I_{s0}/2\pi a = \sqrt{b/a} H_s$, which is much smaller than $H_s$ when $b \ll
a$, the case of interest here.

When the current
$I$ rises slightly above
$I_{s0}$, vortices and antivortices
nucleate on opposite sides of the strip and 
then experience a Lorentz driving force per unit
length of magnitude $f_d(x) = j_z(x) \phi_0$, which is sharply peaked at
the edges and reaches a minimum at the center of the strip [see Eq.\
(\ref{jzMeissner})].  Let us consider the case for which the bulk-pinning current
density
$J_p$ is independent of $B$ (the Bean-London model\cite{Bean62,London63}).  When
$J_p$ is less than
$I_{s0}/2
\pi a b$, the driving forces exceed the bulk-pinning forces for all $x$, and the
nucleated vortices and antivortices are driven towards each other until they
annihilate.  We therefore see that the self-field critical current $I_c$ of the
strip is governed entirely by the edge barrier, i.e., $I_c = I_{s0}$,
independent of the bulk-pinning
$J_p$,  when the ratio $p = I_p/I_{s0}$ is less than $2/\pi$, where  
$I_p = 4ab
J_p$ is defined to be the bulk-pinning critical current in the
absence of the edge barrier.

For larger values of $J_p$, i.e., for $p > 2/\pi$,  
the critical current
$I_c$ of the strip (where a voltage first appears along the length) is larger
than the current
$I_{s0}$ when vortices are first nucleated. As the current $I$ slightly exceeds
$I_{s0}$, the large screening currents near the sample edges drive nucleating
vortices and antivortices toward the center of the strip, but they  stop
where $j_z(x) = J_p$.  As shown in Ref.\ \onlinecite{Maksimov02}, further
increases of the current produce metastable magnetic-flux distributions
initially characterized by five zones.    Near the edges are two
vortex-free zones where the local current density $j_z(x) > J_p$.  Along the
centerline is another vortex-free zone where  $j_z(x) < J_p$.  On the right
side of the central vortex-free zone is a region containing vortices
with
magnetic flux density $B_y(x) > 0$, in which $j_z(x) = J_p$; and
on  the left
side of the central vortex-free zone is a region containing antivortices with
magnetic flux density $B_y(x) < 0$ in which $j_z(x) = J_p.$   The critical
current $I_c$ is reached when $I$ and $j_z(x)$ are just large enough to drive the
vortex-filled and antivortex-filled regions together, shrinking the central
vortex-free zone to zero width, such that  vortex-antivortex annihilation can
occur along the centerline. 

In zero applied field when  $p$ is just above $2/\pi$, the state of the strip
when $I$ is barely above the critical current $I_c$ therefore can be described
qualitatively in terms of four zones as follows:  Vortices and antivortices
nucleate at opposite edges of the strip, where the magnitude of the local
magnetic field is equal to
$H_s$. Once nucleated, vortices (antivortices) move rapidly to the left (right)
through otherwise vortex-free zones near the strip edges where the supercurrent
density
$j_z(x)$ exceeds the bulk zero-field depinning critical current density
$J_p$.  The vortices (antivortices) then enter a vortex-filled zone on the
right (left) side of the centerline, where the magnetic flux density $B_y >
0$ ($B_y < 0$) and the local critical current density
$j_z(x)$ is just barely above the  depinning critical current density
$J_p$.
The critical current $I_c$ is larger than $I_{s0}$ but also
larger than $I_p$ because $j_z(x) > J_p$ in the vortex-free zones.
Analytic
expressions from which 
$I_c$ can be calculated  are given in Refs.\ \onlinecite{Maksimov02} and
\onlinecite{Elistratov02} for the case that $J_p$ is
independent of
$B$.

For increasing values of $J_p$ or $p = I_p/I_{s0}$, the
vortex-free zones at $I = I_c$ become narrower, and they shrink to zero width
when the magnitude of the $J_p$-generated self-field at the sample
edges becomes equal to $H_s$.  If $J_p$ is independent of $B$ and $b
\ll a$, the self-field at $(x,y) = (\pm a,0)$ can be shown from Eq.\ (\ref{Byr})
to be
\begin{equation}
H_y(\pm a,0) = \pm (J_p b/
\pi)[\ln (2a/b) + 1]
\label{Hya}
\end{equation}with correction terms inside the brackets of order
$(b/a)^2$.  The value of $J_p$ that makes $H_y(a,0) = H_s$ is 
$J_{ps} = \pi H_s/b[\ln (2a/b)+1],$ and the corresponding value of $p$ is 
\begin{equation}
p_s =
\frac{I_{ps}}{I_{s0}}= \frac{2\sqrt{a/b}}{\ln (2a/b)+1},
\label{ps}
\end{equation} where $I_{ps}= 4abJ_{ps}$.  

For larger values of the depinning critical current density, $J_p > J_{ps}$
or $p \ge p_s$, the critical current is given by $I_c = I_p$ to excellent
approximation because
$J_z(x) = J_p$ throughout the entire width of the strip except very near the
edges.  Any surface-current contribution at the edges arising from a
surface-barrier-induced discontinuity in $H$ can be shown to be smaller
than
$I_c$ by a factor of order $b/a$.

Such complicated magnetic-field and current-density distributions arising from
the geometrical edge barrier are important for strips fabricated of low-pinning
superconductors, such as Bi-2212.  On the other hand, for state-of-the-art YBCO
films and prototype second-generation coated conductors, the depinning critical
current density is so high that the parameter
$p$ is well above $p_s$, and the excess current due to the edge barrier is
negligible.   For example, Rupich et al.\cite{Rupich03} recently reported
measurements of a self-field critical current of $I_c = 112$ A at 77 K for a
coated conductor composite of length 1.25 m and width 1 cm, in
which the superconductor was metal-organic-derived YBCO of thickness 0.8
$\mu$m coated on a deformation-textured NiW alloy substrate buffered with
Y$_2$O$_3$/YSZ/CeO$_2$.  We can estimate $I_{s0}$ as follows.  The
authors of Ref.\
\onlinecite{Hao91} found from
magnetization measurements, using a $1-(T/T_c)^2$ temperature
dependence for $H_c(T)$, $\sqrt 2 H_c(0) = 1.4 \times 10^4$ Oe, $\kappa =
57$, and $T_c = 93.9$ K for a YBCO
single crystal, which yields $H_c$ = 3.24 kOe = 2.6 $\times 10^5$ A/m at $T = 77$
K.  Taking the nucleation field $H_s$ to be the same as $H_{c1}$ and using the
Ginzburg-Landau expression
$H_{c1} = (H_c/\kappa
\sqrt 2) (\ln
\kappa + 0.5)$, we obtain $H_s = H_{c1} = 180 $ Oe $ = 1.5 \times 10^4$ A/m
and $I_{s0} = 4.1$ A.  Taking  $I_p = I_c$ = 112 A, we obtain $p =
I_p/I_{s0} = 27$, such that $p  \gg p_s$, where $p_s = 3.1$ 
[Eq.\ (\ref{ps})].  This indicates that the vortex-filled region in 
which the current density
is at the depinning critical current density fills practically the entire cross
section at the critical current of coated
conductors such as that reported in Ref.\
\onlinecite{Rupich03}.

\section{Bulk pinning} 

For the remainder of this paper, we limit our attention to materials in which
the depinning critical current density is so high that geometrical edge barriers
have a negligible effect upon the critical current.  We now focus on
the questions of how to calculate the the average critical current density
$J_c = I_c/4ab$ in an applied magnetic induction ${\bm B}_a$ and how to determine
the magnetic-field and current-density distributions at
$J_c$ when the local depinning critical current density $J_p$ has a significant
dependence upon the local magnetic induction $\bm B$.

\subsection{Basic equations}

Consider a type-II superconducting strip with a geometry as sketched in Fig.\ 1.
If the current density associated with the magnetic induction $\bm{B = \nabla
\times A}$ is $\bm{j}(x,y) = (1/\mu_0)\bm{\nabla \times B}(x,y)$, the vector
potential $\bm{A}(x,y) = \hat{z} A_z(x,y)$ generated by $\bm{j}(x,y)$ can be
calculated from 
\begin{multline}
A_{zr}(x,y,a,b)=  \\ 
-\frac{\mu_0}{4 \pi}\int_{-a}^a  \int_{-b}^b 
 j_z(u,v)\ln[(x-u)^2+(y-v)^2]dudv,
\label{Azrgen}
\end{multline}
where the subscript $r$ refers to the
rectangular cross section.
If $\bm{j} = \hat{z}j_z$ and the sheet current density $K_z = 2bj_z$ are
constant over the cross section of the strip, the resulting vector potential and
the
$x$ and
$y$ components of the magnetic induction are  
\begin{align}
A_{zr}(x,y,a,b)&= (\mu_0 K_z/2 \pi)a_{zr}(x,y,a,b),
\label{Azr}\\
B_{xr}(x,y,a,b)&= (\mu_0 K_z/2 \pi)b_{xr}(x,y,a,b),
\label{Bxr}\\
B_{yr}(x,y,a,b)&= (\mu_0 K_z/2 \pi)b_{yr}(x,y,a,b),
\label{Byr}
\end{align}
where expressions for the functions
$a_{zr},$
$b_{xr},$ 
$b_{yr},$ and their partial derivatives are given in Appendix A.
Shown in Fig.\ 1 are contours of constant $A_{zr}(x,y)$, which correspond to
${\bm B}_r$ field lines.

In the limit of vanishing thickness of the strip of width $2a$, simpler results
but with singularities at the edges can be obtained by replacing $j_z$ in Eq.\
(\ref{Azrgen}) by $j_z(x,y) = K_z(x) \delta(y)$, such that the vector potential
becomes
\begin{equation}
A_{zs}(x,y,a)=  \\ 
-\frac{\mu_0}{4 \pi}\int_{-a}^a 
 K_z(u)\ln[(x-u)^2+y^2]du,
\label{Azsgen}
\end{equation}
where the subscript $s$ is a reminder that the results apply to a strip in the
limit of zero thickness.
If $K_z$ is uniform over the width of the strip, the resulting vector
potential and the
$x$ and
$y$ components of the magnetic induction are  
\begin{align}
A_{zs}(x,y,a)&= (\mu_0 K_z/2 \pi)a_{zs}(x,y,a),
\label{Azs}\\
B_{xs}(x,y,a)&= (\mu_0 K_z/2 \pi)b_{xs}(x,y,a),
\label{Bxs}\\
B_{ys}(x,y,a)&= (\mu_0 K_z/2 \pi)b_{ys}(x,y,a),
\label{Bys}
\end{align}
where expressions for the functions
$a_{zs},$
$b_{xs},$ and
$b_{ys}$ are given in Appendix B.

We are interested in calculating the critical current $I_c$ as a function of
the applied field when the magnetic induction ${\bm B}_a$ is applied
perpendicular to the $z$ axis.  If one equates the current
density
$j_z$  to the depinning critical current density $J_p({\bm B})$ in Eq.\
(\ref{Azrgen}), which determines the vector potential $\bm A$ describing the
self-fields, one must  self-consistently determine the spatial variation of the
local magnetic induction
${\bm B} = {\bm B}_a + {\bm
\nabla \times \bm A}$.  In general, $J_p({\bm B})$ is not
constant over the cross section of the sample, and the critical current density
$J_c({\bm B}_a) = I_c/4ab$ is therefore not equal to  $J_p({\bm B}_a)$. 

\subsection{General procedure with a 2D grid and $J_p(\bm B)$}

The following general numerical procedure, which is similar to that used by
previous authors\cite{Daeumling89,Conner91} for the disk geometry,
could be used to account for self-field effects when the depinning critical
current density $J_p(\bm B)$ depends upon both the magnitude and direction of
the local magnetic induction $\bm B$.  Divide the rectangular cross
section into
$N = N_xN_y$ current elements of dimensions
$\Delta x
\times \Delta y$, where $\Delta x = 2a/N_x$ and $\Delta y = 2b/N_y$, centered at
the coordinates $(x_n,y_n)$, where $n = 1, 2, ..., N$.  If $N$ is large, the
bulk-pinning critical current density $J_p({\bm B}_n)$ in element $n$ is very
nearly constant. The local magnetic induction in element $n$ can be
expressed as
${\bm B}_n = {\bm B}(x_n,y_n)= {\bm B}_a + {\bm B}_{self}(x_n,y_n),$
 where  ${\bm
B}_{self}(x_n,y_n)= {\hat x} B_{self,x}(x_n,y_n)+{\hat y} B_{self,y}(x_n,y_n) $
can be obtained with the help of Eqs.\ (\ref{Bxr}) and (\ref{Byr}) by summing the
contributions from all current elements: 
\begin{align}
B_{self,x}(x,y)&= \sum_{m=1}^N\frac{\mu_0 K_m}{2 \pi}
b_{xr}(x-x_m,y-y_m,\frac{\Delta x}{2},\frac{\Delta y}{2}),
\label{Bselfx}\\
B_{self,y}(x,y)&= \sum_{m=1}^N\frac{\mu_0 K_m}{2 \pi}
b_{yr}(x-x_m,y-y_m,\frac{\Delta x}{2},\frac{\Delta y}{2})
\label{Bselfy}
\end{align}
with $K_m = J_p({\bm B}_m)\Delta y$.
This results in $2N$ equations, which for each given bulk-pinning
critical current density function $J_p({\bm B})$ can be solved by iterative
methods to obtain the
$x$ and $y$ components of the $N$ vectors ${\bm B}_n$.  The average critical
current density can then be calculated as follows:
\begin{equation}
J_c({\bm B}_a) = \frac{1}{N}\sum_{n=1}^N J_p({\bm B}_n).
\label{Jcavg}
\end{equation}

\subsection{General procedure for strips with a 1D grid and $J_p(B_x,B_y)$}

Suppose the strip is subjected to a uniform applied magnetic induction ${\bm
B}_a = \hat x B_{ax} + {\hat y} B_{ay}$.  We consider the case in which the
underlying depinning critical current density
$J_p(B_x,B_y)$ is a known function of both the $x$ and $y$ components of the
local magnetic induction $\bm B$. For example, $J_p(B_x,B_y)$ could be determined
experimentally by (a) measuring $J_c$ in applied fields ${\bm B}_a$ large
enough that the effects of the self-field $\bm B_{self}$ are
negligible and (b) fitting to a model that allows extrapolation to small $B$. 
In calculating the remanent magnetization of disks in
self-fields, the authors of Refs.\
\onlinecite{Daeumling89} and
\onlinecite{Conner91} found for the disk geometry that the magnitude of $\bm B$
changed so significantly over the disk thickness that the depinning
critical current density at the top and bottom surfaces was significantly
suppressed relative to that midway between the two surfaces. 
This suppression of the depinning current density near  the top and bottom
surfaces must also be taken into account to achieve self-consistent solutions
accounting for self-field effects upon the  average critical current density
$J_c(\bm B_a)$ vs $\bm B_a$ of long strips.  

There are three properties of thin
superconducting strips with
$b \ll a$ (see Fig.\ 1) that enable us to accomplish this  using only a 1D
grid by  applying an idea
from Refs.\ \onlinecite{Conner91} and \onlinecite{Mikitik00}.  We simplify
the general procedure  outlined in Sec.\ III B by taking
$N_y = 1$, such that
$N = N_x$, and by  considering current elements of dimensions $\Delta x = 2a/N$
and
$\Delta y = 2b$, centered at the coordinates $(x_n,0)$, where $n = 1, 2, ...,
N$.
For simplicity, we use the approximation that $\bm B \approx \mu_0 \bm H$, such
that $\bm J \approx \bm j = (1/\mu_0)\bm{\nabla \times B}$.

The first important  property of thin strips we use is that
the
$x$ and
$y$ components of 
$\bm B_{self}(x,y)$ can be obtained to good accuracy by writing
\begin{align}
B_{self,x}(x,y)&= \sum_{m=1}^N\frac{\mu_0 K_m}{2 \pi}
b_{xr}(x-x_m,y,\frac{\Delta x}{2},b),
\label{Bselfx1D}\\
B_{self,y}(x,y)&= \sum_{m=1}^N\frac{\mu_0 K_m}{2 \pi}
b_{yr}(x-x_m,y,\frac{\Delta x}{2},b),
\label{Bselfy1D}
\end{align}
where $K_n = 2b\bar j_z(x_n)$ is the average sheet current density in the
element at $x = x_n$.  The net magnetic induction is $\bm B = \bm B_a + \bm
B_{self}$, where 
\begin{eqnarray}
B_x(x,y)= B_{ax} + B_{self,x}(x,y),\\
\label{Bx1}
B_y(x,y)= B_{ay} + B_{self,y}(x,y).
\label{By1D}
\end{eqnarray}
As can be seen from Fig.\ 1, $B_{self,x}(x,-b) > 0$ and $B_{self,x}(x,b)
=-B_{self,x}(x,-b)$.

The second property of thin films we use is that, although  the
current density $\bm{j}(x,y) = j_z(x,y)
\hat{z}$ is given in general by $j_z =(1/\mu_0)
(\partial B_y/\partial x-\partial B_x/\partial y),$ it is a good approximation
for most of the strip width to write $j_z = - (1/\mu_0)
\partial B_x/\partial y$ when $b \ll a$. 
The accuracy of this approximation can be estimated  
for a uniform current
density in a strip of rectangular cross section
by calculating the ratio
$R_J(x,y) = (\partial B_y/\partial x)/(-\partial B_x/\partial y)$ using Eqs.\
(A6) and (A7).  Although $R_J \sim 1$ at the edges of the strip, we find that
$R_J(0,0)  \approx
(2/\pi)(b/a)$ at the center of the strip when $b/a \ll 1$.
When $b/a = 0.01$, $R_J(0,0) = 0.0064$ 
and $R_J(x,0) < 0.1$ for $a-|x| > 3.5 b$.
When $b/a = 0.001$, $R_J(0,0) = 0.00064$ 
and  $R_J(x,0) < 0.1$ for $a-|x| > 3.5 b$.

The third property of thin films we use is that 
$  |\partial B_y/\partial y| \ll |\partial B_x/\partial y|$, such that although
$B_x(x,y)$ changes considerably over the film thickness, $B_y(x,y)$ is very
nearly a constant, equal to $B_y(x,0)$.  
The accuracy of this approximation can be estimated  
for a strip carrying a uniform current
density
by  calculating the ratio
$R_B(x) = |[B_y(x,b)-B_y(x,0)]/B_y(x,0)]$ using Eq.\ (A3).
Although $R_B(x)$ is largest near the edges of the strip, $R_B(x) < 0.01$ for  
$a-|x| > 3.5 b$ when $b/a \le 0.01$.

The second property allows us to obtain the average of
$j_z(x,y)$ over the sample thickness from
$\bar j_z(x) = B_{self,x}(x,-b)/\mu_0b$, and the third property allows us to set
$B_y(x,y) = B_y(x,0)$ in the critical state force-balance
expression
\begin{equation}
j_z(x,y) =J_p(B_x(x,y),B_y(x,y)),
\label{jbalance}
\end{equation}
such that in Eqs.\ (\ref{Bselfx1D}) and (\ref{Bselfy1D}) we may write 
\begin{eqnarray}
K_n = 2b \bar j_z(x_n) = 2B_{self,x}(x_n,-b)/\mu_0,
\label{Kn}
\end{eqnarray}
where $B_{self,x}(x_n,-b)$ is obtained either numerically or analytically from 
(see  Refs.\ \onlinecite{Conner91} and \onlinecite{Mikitik00})
\begin{equation}
\int_{B_{ax} - B_{self,x}(x_n,-b)}^{B_{ax} + B_{self,x}(x_n,-b)}
\frac{du}{J_p(u,B_y(x_n,0))} 
= 2\mu_0 b.
\label{Bxint}
\end{equation}

One may now use the above equations to find self-consistent solutions for $\bm
B$ and $\bm j$.  The $2N$ unknowns, $B_x(x_n,-b)$  and
$B_y(x_n,0),$ can be obtained numerically from $2N$ equations obtained from
Eqs.\ (18-20),  (\ref{Kn}), and
(\ref{Bxint}).  Once these solutions have been found, we may calculate the
average critical current density from  
\begin{eqnarray}
J_c({B}_a) = \frac{1}{N}\sum_{n=1}^N \bar j_z(x_n)
=\frac{1}{N\mu_0 b}\sum_{n=1}^N B_{self,x}(x_n,-b).
\label{Jcavg1D}
\end{eqnarray}

\subsection{Calculations in a perpendicular field with a 1D grid and a Kim-model
$J_p(B(x,y))$}

We have applied the numerical procedure described in Sec.\ III C  
to calculate $J_c(B_a)$ in a perpendicular field $\bm B_a = \hat y B_a$
for three ficticious YBCO samples (a, b, and c) with dimensions as in Fig.\ 1 but
with 
$a$ = 120 $\mu$m and $b$ = (a)  0.1 $\mu$m, (b) 0.5 $\mu$m, and (c) 1.5
$\mu$m.  Suenaga et al.\  \cite{Suenaga04} reported that the depinning critical
current densities inferred from 77 K magnetization measurements on three YBCO
disks with these thicknesses were well fit by the Kim model,\cite{Kim63}
\begin{equation}
J_{pK}(B) = J_{pK}(0) /(1+|B|/B_0),
\label{JpK}
\end{equation} 
except at low fields where
self-field effects are responsible for deviations.
For $J_p(B_x,B_y)$ in Eq.\ (\ref{Bxint}) we therefore took
the local depinning critical current density to be $J_p(B_x,B_y) = J_{pK}(B)$,
where $B = \sqrt{B_x^2 +B_y^2}$. The parameters were
(a)
$J_{pK}(0)$ = 4.32  $\times 10^6$ A/cm$^2$ and
$B_0$ = 17.6 mT, (b) $J_{pK}(0)$ = 3.22  $\times 10^6$ A/cm$^2$ and $B_0$
= 18.8 mT, and (c) $J_{pK}(0)$ = 2.55  $\times 10^6$ A/cm$^2$ and $B_0$
= 22.6 mT, respectively.\cite{Suenaga04}  With Eq.\ (\ref{JpK}) giving the
explicit form for $J_p$, it was possible to evaluate Eq.\ (\ref{Bxint})
analytically with the result
\begin{flalign}
&2B_0B_x(x_n,-b)+B_x(x_n,-b)\sqrt{B_x^2(x_n,-b)+B_y^2(x_n,0)} \nonumber \\
&+B_y^2(x_n,0)\sinh^{-1}[B_x(x_n,-b)/|B_y(x_n,0)|] \nonumber \\
&=2\mu_0J_{pK}(0)B_0b.
\label{Bx2} 
\end{flalign}

Using Mathematica,\cite{Math} we  solved Eqs.\  (18-20),  (\ref{Kn}), 
and (\ref{Bx2}) numerically with $N$ = 101 to obtain  $B_x(x_n,-b)$  and
$B_y(x_n,0)$ for samples a, b, and c.
This allowed us to calculate $B_x(x,y)$ and $B_y(x,y)$ from  Eqs.\
(\ref{Bselfx1D}), (\ref{Bselfy1D}), (\ref{Kn}), and
\begin{flalign}
B_x(x,y) &= B_{self,x}(x,y),
\label{Bx} \\
B_y(x,y) &= B_a + B_{self,y}(x,y).
\label{By}
\end{flalign}  
We also calculated $B_a^*$, the value of $B_a$ at which
$B_y(-a,0) = 0$, for the three
samples considered in this paper:  
$B_a^*/B_0$ = (a) 0.651, (b) 1.320, and (c) 1.834 or $B_a^*$ = (a) 11.5 mT,
(b) 24.8 mT, and (c) 41.4 mT.
Shown in Fig.\ 2 are values of $J_c(B_a)$ vs $B_a$  calculated from Eq.\
(\ref{Jcavg1D}). When $B_a \ge B_a^*$, the calculated
values of
$J_c(B_a)$ agreed with $J_{pK}(B_a)$ within (a) 0.6\%, (b) 1.3\%, and (c) 2.1\%
for the three samples, where the values of $B_a^*$ are shown by the solid
symbols.
Note that the suppression of $J_c(0)$ is greatest for the thickest film.  
In general, self-field effects become more important as the strip thickness
increases.

\begin{figure}
\includegraphics[width=8cm]{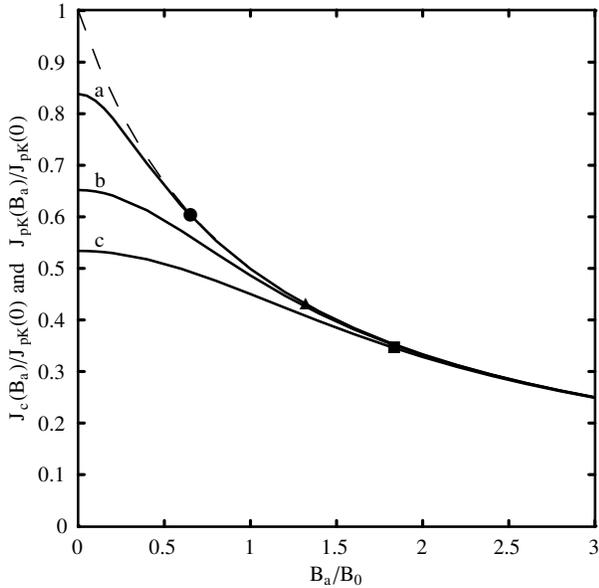}
\caption{The normalized average critical current density $J_c(B_a)/J_{pK}(0)$ vs
$B_a/B_0$ calculated from Eq.\ (\ref{Jcavg1D}) for the
three samples a, b, and c (solid), compared with the Kim model,
$J_{pK}(B_a)/J_{pK}(0)$, Eq.\ (\ref{JpK}) (dashed).  The values of
$B_a^*/B_0$ for the three samples are identified by the (a) solid circle, (b)
solid triangle, and (c) solid square.}
\label{Fig2}
\end{figure}

Shown in Fig.\ 3 are  normalized plots of
$B_y(x,0)$ vs
$x/a$ for the thickest of the three samples for applied fields $B_a = $ (a)
0, (b)
$B_a^* /2$, (c) $B_a^*$, and (d) $3B_a^* /2$.  Note that $B_y = 0$ at the
left edge of the sample when $B_a = B_a^*$.  Shown in Fig.\
4 are corresponding normalized plots of 
\begin{equation}
\bar j_z(x) = B_x(x,-b)/\mu_0 b
\label{jzbar}
\end{equation}
 vs
$x/a$.  Note that $\bar j_z(x)$ is largest  at those values of $x$ where
$B_y(x,0) = 0$ in Fig.\ 3. 
Note also that when $B_a > B_a^*$, $\bar j_z$ vs $x$ becomes flatter; this is
another indication that self-field effects have less of an effect upon $J_c(B_a)$
as
$B_a$ increases.

A rough (within about a factor of 1.2) estimate of  $B_a^*$, perhaps useful
in the analysis of  experiments, is
$\mu_0 H_y(a,0)$, where
$J_p$ in Eq.\ (\ref{Hya}) is replaced by the experimental value of $J_c(0)$.
Using the solid curves in Fig.\ 2 to obtain such an estimate yields (a) 12.7 mT,
(b) 30.1 mT, and (c) 49.6 mT for the three samples discussed above, for which
$J_c(0)/J_{pK}(0)$ = (a) 0.838, (b) 0.652, and (c) 0.534, or $J_c(0)$ = (a)
3.62  $\times 10^6$ A/cm$^2$, (b) 2.10  $\times 10^6$ A/cm$^2$, and (c)
1.36  $\times 10^6$ A/cm$^2$, respectively.

\begin{figure}
\includegraphics[width=8cm]{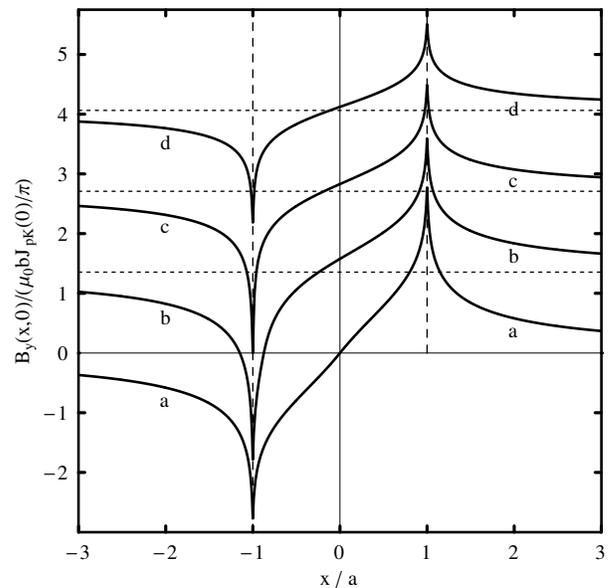}
\caption{Normalized magnetic flux density $B_y(x,0)$ vs
$x/a$ for the thickest of the three samples, calculated
from Eq.\ (\ref{By}), for applied fields 
$B_a = $ (a)
0, (b)
$B_a^* /2$, (c) $B_a^*$, and (d) $3B_a^* /2$ (applied fields
shown as  horizontal dashed lines), where $B_a^*$ = 41.4 mT and $B_a^*/(\mu_0
b J_{pK}(0)/\pi) = 2.71.$}
\label{Fig3}
\end{figure}

\begin{figure}
\includegraphics[width=8cm]{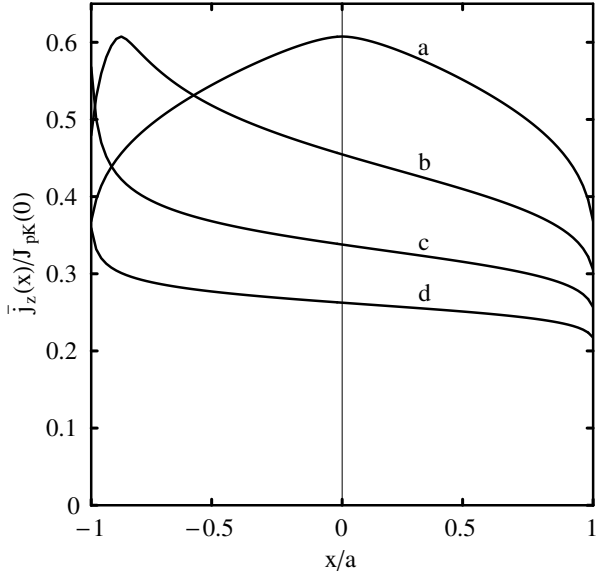}
\caption{Local current density averaged over the sample thickness
$\bar j_z(x)/J_{pK}(0)$ vs
$x/a$ for the thickest sample, calculated from Eq.\ (\ref{jzbar}) for applied
fields 
$B_a = $ (a)
0, (b)
$B_a^* /2$, (c) $B_a^*$, and (d) $3B_a^* /2$, where $B_a^*$ = 41.4 mT.}
\label{Fig4}
\end{figure}

  Several features of the above behavior of 
$J_c$ vs
$B_a$ using the Kim model\cite{Kim63} are close to those found
analytically for
$J_c(B_a)$ when the applied field is parallel to an infinite slab, as discussed
in Appendix C. 

\section{Summary}

In Sec.\ II, we briefly reviewed the behavior of the average critical current
density $J_c(B_a)$ of superconducting strips when both bulk and edge pinning
play a role.  The main conclusion of this section is that edge pinning has a
negligible effect upon $J_c$ in state-of-the-art YBCO  films
and prototype second-generation coated conductors.

In Sec.\ III, we considered the problem of how to account self-consistently for
the dependence of the local depinning critical current density upon the local
magnetic induction when calculating $J_c(\bm B_a)$.  In Sec. III A, we presented
some basic equations, and in Sec.\ III B, we set up, but did not apply, a
procedure that, using a 2D grid and starting from an underlying local depinning
critical current density
$J_p(B_x,B_y)$, would yield the average critical current density
$J_c(\bm B_a) = I_c/4ab$ vs an applied magnetic induction $\bm B_a$ for a sample
of rectangular  cross section of arbitrary width $2a$ and thickness $2b$.

In Sec.\ III C, we developed a general
method  that, using a 1D grid and starting from an underlying local depinning
critical current density
$J_p(B_x,B_y)$, yields the average critical current density
$J_c(\bm B_a) = I_c/4ab$ as a function of the magnetic induction $\bm B_a$
applied to a flat sample whose width $2a$ is much larger than its thickness
$2b$, a case of considerable interest to those involved in research on prototype
second-generation YBCO coated conductors.  This method should work well even
when $J_p$ is a function of both the magnitude
$B$ and the direction $\hat B$ of the local magnetic induction $\bm B = B
\hat B$. 

In Sec.\ III D, we applied the above method to
calculate $J_c(B_a)$ in a perpendicular field $B_a$ using the Kim
model\cite{Kim63} [Eq.\ (\ref{JpK})], in which  the local depinning critical
current density
$J_p$ is a function of the magnitude
$B$ but not the direction $\hat B$ of  $\bm B = B
\hat B$. 
The functional form of the
$B$ dependence of the Kim model's $J_p(B)$ resulted in the particular form
of Eq.\ (\ref{Bx2}), which we used in determining $B_x(x,-b)$ and $\bar j_z(x)$.
However, our general method is not limited to the Kim model but could be used
with any other realistic model (whether analytic or numerical) for  
$J_p(B).$ Instead of using Eq.\ (\ref{Bx2}), 
one could use Eq.\ (\ref{Bxint}) to obtain a new algorithm connecting
$B_x(x_n,-b)$ and
$B_y(x_n,0)$ from which $J_c(B_a)$, $B_y(x,0)$, and $\bar j_z(x)$ could
be determined.

In this paper we have considered only static current and field distributions at
the critical current.  For extensive treatments of the dynamics of current and
field penetration into strips, see
Refs. \onlinecite{Mikitik00,Brandt94a,
Brandt96,Brandt01}.
 
\begin{acknowledgments}
We thank E. H. Brandt, S. Foltyn, V. G. Kogan, M. P. Maley, and M. Suenaga for
stimulating discussions.  This manuscript has been authored in part by Iowa
State University of Science  and Technology under Contract No.\ W-7405-ENG-82
with the U.S.\ Department of  Energy.
\end{acknowledgments}

\appendix

\section{$A_z$, $B_x$, and $B_y$ for a strip carrying
uniform sheet current density}

When a  strip of rectangular cross section (width $2a$ and thickness $2b$)
carries a uniform sheet current density
$K_z = 2b j_z$ as described in Sec.\ I, the vector potential derived from Eq.\
(\ref{Azrgen}) (neglecting an unimportant additive constant) can be expressed as
in Eq.\ (\ref{Azr}), where 
\begin{flalign}
&a_{zr}(x,y,a,b)=\nonumber \\
&\{(x-a)^2[\arctan(\frac{y+b}{x-a})-\arctan(\frac{y-b}{x-a})]\nonumber\\
&+(x+a)^2[\arctan(\frac{y-b}{x+a})-\arctan(\frac{y+b}{x+a})]\nonumber\\
&+(y-b)^2[\arctan(\frac{x+a}{y-b})-\arctan(\frac{x-a}{y-b})]\nonumber\\
&+(y+b)^2[\arctan(\frac{x-a}{y+b})-\arctan(\frac{x+a}{y+b})]\nonumber\\
&+(x-a)(y+b)\ln[(x-a)^2+(y+b)^2]\nonumber\\
&-(x-a)(y-b)\ln[(x-a)^2+(y-b)^2]\nonumber\\
&+(x+a)(y-b)\ln[(x+a)^2+(y-b)^2]\nonumber\\
&-(x+a)(y+b)\ln[(x+a)^2+(y+b)^2]\}/4b.
\label{azr}
\end{flalign}
The $x$ and $y$ components of the magnetic induction derived from 
Eq.\ (\ref{Azr}) and
$\bm{B = \nabla \times A}$ are given in Eqs.\ (\ref{Bxr}) and (\ref{Byr}), where 
\begin{flalign} &b_{xr}(x,y,a,b)= \nonumber \\
&\space\frac{y-b}{2b}[\arctan(\frac{x+a}{y-b})-\arctan(\frac{x-a}{y-b})]
\nonumber\\
&+\frac{y+b}{2b}[\arctan(\frac{x-a}{y+b})-\arctan(\frac{x+a}{y+b})]
\nonumber\\
&+\frac{x+a}{4b}\ln[\frac{(x+a)^2+(y-b)^2}{(x+a)^2+(y+b)^2}]
\nonumber\\
&+\frac{x-a}{4b}\ln[\frac{(x-a)^2+(y+b)^2}{(x-a)^2+(y-b)^2}],
\label{bxr}
\end{flalign}
\begin{flalign}
&b_{yr}(x,y,a,b)=\nonumber \\
&\frac{x-a}{2b}[\arctan(\frac{y-b}{x-a})-\arctan(\frac{y+b}{x-a})]\nonumber\\
&+\frac{x+a}{2b}[\arctan(\frac{y+b}{x+a})-\arctan(\frac{y-b}{x+a})]\nonumber\\
&+\frac{y-b}{4b}\ln[\frac{(x-a)^2+(y-b)^2}{(x+a)^2+(y-b)^2}]\nonumber\\
&+\frac{y+b}{4b}\ln[\frac{(x+a)^2+(y+b)^2}{(x-a)^2+(y+b)^2}].
\label{byr}
\end{flalign}

Partial derivatives of the components of $\bm B$, such as $\partial
B_{xr}/\partial y = (\mu_0 K_z/2\pi)\partial b_{xr}/\partial y $, are given by
the following equations,
\begin{flalign} 
&\partial b_{xr}(x,y,a,b)/\partial x = \nonumber \\
&\ln[\frac{(x+a)^2+(y-b)^2}{(x+a)^2+(y+b)^2}
\frac{(x-a)^2+(y+b)^2}{(x-a)^2+(y-b)^2}]/4b,\\
\label{bxxr}
&\partial b_{yr}(x,y,a,b)/\partial y 
= -\partial b_{xr}(x,y,a,b)/\partial x,\\
\label{byyr}
&\partial b_{xr}(x,y,a,b)/\partial y = \nonumber \\
&[\arctan(\frac{x+a}{y-b})+\arctan(\frac{x-a}{y+b}) \nonumber \\
&-\arctan(\frac{x+a}{y+b})-\arctan(\frac{x-a}{y-b})]/2b,\\																																										
\label{bxyr} 
&\partial b_{yr}(x,y,a,b)/\partial x = \nonumber \\
&[\arctan(\frac{y-b}{x-a})+\arctan(\frac{y+b}{x+a}) \nonumber \\
&-\arctan(\frac{y+b}{x-a})-\arctan(\frac{y-b}{x+a})]/2b.																																										
\end{flalign}

\section{$A_z$, $B_x$, and $B_y$  for a thin strip carrying
uniform sheet current density}

When a thin strip of width $2a$ and negligible thickness carries a uniform
sheet current density
$K_z$ as described in Sec.\ I, the vector potential derived from Eq.\
(\ref{Azsgen}) (neglecting an unimportant additive constant) can be expressed as
in Eq.\ (\ref{Azs}), where 
\begin{multline}
a_{zs}(x,y,a)= 
y[\arctan(\frac{x-a}{y})-\arctan(\frac{x+a}{y})]\\
+\frac{x-a}{2}\ln[(x-a)^2+y^2]
-\frac{x+a}{2}\ln[(x+a)^2+y^2].
\label{azs}
\end{multline}
The $x$ and $y$ components of the magnetic induction derived from 
Eq.\ (\ref{Azs}) and
$\bm{B = \nabla \times A}$ are given in Eqs.\ (\ref{Bxs}) and (\ref{Bys}), where 
\begin{equation}
b_{xs}(x,y,a)= 
\arctan(\frac{x-a}{y})-\arctan(\frac{x+a}{y})],
\label{bxs}
\end{equation}
and
\begin{equation}
b_{ys}(x,y,a)= 
\frac{1}{2}\ln[\frac{(x+a)^2+y^2}{(x-a)^2+y^2}].
\label{bys}
\end{equation}

\section{$J_c(B_a)$ for $B_a$ parallel to a slab} 

Although our main interest in this paper is in the average critical current
density  $J_c(B_a)$ vs $B_a$ when a magnetic field is applied
perpendicular to a flat strip, for comparison we present here analytic results
for the average critical current
density  $J_c(B_a)$ in the $z$ direction when a  magnetic
induction $B_a$ is applied in the
$y$ direction parallel to the surface of an infinite  slab of thickness $2a$.  We
consider two models for the depinning critical current density
$J_p(B)$ and assume that  $B =
\mu_0 H$, such that at the critical current we have for each model 
\begin{equation}
J_z(x) = J_p(B_y) = \frac{1}{\mu_0} \frac{dB_y(x)}{dx}.
\label{Jzslab}
\end{equation}
Associated with this current is the self-field $B_s$, such that $B_y(\pm a) =
B_a \pm B_s$.  Integrating Eq.\ (\ref{Jzslab}) from one side of the slab to the
other, we find that the average critical current density is
\begin{equation}
J_c = \frac{B_s}{\mu_0 a},
\label{Jcslab}
\end{equation}
where the dependence of $B_s$ upon $B_a$ can be determined by solving
Eq.\ (\ref{Jzslab}) for $B_y(x)$. For both models of $J_p(B)$, we find that
$J_c(B_a)$  has an inflection point at $B_a=B_a^*$ and is equal to
$J_p(B_a)$ for
$|B_a|  \ge B_a^*$.

\subsection{$J_p(B) = \alpha/|B|$}

For the model $J_p(B) = \alpha/|B|$, we find that $B_a^* = \sqrt{\mu_0
\alpha a}$ and 
\begin{eqnarray}
\frac{J_c(B_a)}{J_c(B_a^*)} & =& \sqrt{2-(\frac{B_a}{B_a^*})^2}, 
0 \le |B_a| \le B_a^*,
\\ &= &\frac{B_a^*}{|B_a|}, |B_a| \ge B_a^*. 
\label{Jcalpha}
\end{eqnarray}

\subsection{$J_p(B) = J_{p0}/(1+|B|/B_0)$}

For the Kim model\cite{Kim63}, $J_p(B) =  J_{p0}/(1+|B|/B_0)$,  we find that
\begin{equation}
B_a^* =(B_0/2)
(\sqrt{1+4\mu_0 J_{p0}a/B_0} - 1)
\label{Ba*Kim}
\end{equation} and 
\begin{eqnarray}
\frac{J_c(B_a)}{J_{p0}}& =& \frac{\sqrt{1+2b^*+2b^{*2}-b^2}-1}{b^*(1+b^*)}, 
0 \le b\le b^*,
\\ &= &\frac{1}{1+b}, b \ge b^*, 
\label{JcKim}
\end{eqnarray}
where $b=|B_a|/B_0$ and $b^* = B_a^*/B_0$.



\begin{thebibliography}{99}
\bibitem{Comment} Different expressions for $I_{s0}$ were obtained in
Refs.\ \onlinecite{Maksimov02} and \onlinecite{Elistratov02}, which
considered thin films for which $b <
\lambda$ and the two-dimensional screening length $\Lambda = \lambda^2/b$ plays
an important role.
\end{thebibliography}
\end{document}